\documentclass[aps,prl,twocolumn,showpacs,floatfix]{revtex4}

\usepackage{graphicx}
\usepackage{epsfig}
\usepackage{bm}
\usepackage{latexsym}

\begin{document}

\title{Metastability and uniqueness of vortex states at depinning}

\author{Mahesh Chandran}
\email{chandran@physics.ucdavis.edu}
\author{R. T. Scalettar}
\author{G. T. Zim\'{a}nyi}
\affiliation{Department of Physics, University of California, Davis, California 95616 }

\date{\today}

\begin{abstract}

We present results from numerical simulations of transport of vortices in the zero-field
cooled (ZFC) and the field-cooled (FC) state of a type-II superconductor. In the absence of
an applied current $I$, we find that the FC state has a lower defect density than the ZFC
state, and is stable against thermal cycling. On the other hand, by cycling $I$,
surprisingly we find that the ZFC state is the stable state. The FC state is metastable as
manifested by increasing $I$ to the depinning current $I_{c}$, in which case the FC state 
evolves into the ZFC state. We also find that all configurations acquire a unique defect 
density at the depinning transition independent of the history of the initial states.

\end{abstract}


\maketitle

Metastability is a generic feature in many disordered systems. Metastable states show
characteristics distinct from the ground state, and are evidenced, for example, by the
path or history dependence of the different response functions. A striking example of
metastability is the difference between the field-cooled (FC) and zero-field cooled
(ZFC) magnetisation in vortex phase of type-II superconductors\cite{bednorz}.  In
superconductors, the FC magnetisation is much smaller than the ZFC magnetisation. This
difference decreases with increasing temperature $T$ and vanishes above the
irreversibility temperature $T_{irr}$. The ZFC state is believed to be metastable and
the FC state to be a quasi-equilibrium state because it is possible to reach the FC
state from the ZFC state by changing $T$ from a low value to $T_{irr}$ and back, but the
ZFC state cannot be reached from a FC state through such a temperature cycle. Similar
features are also observed in spin glasses\cite{binder}. The difference in FC and ZFC
state magnetisation is explained by the emergence of energy barriers below $T_{irr}$,
which inhibits the ZFC state from exploring the full configuration space.

In recent years, transport experiements have shown conflicting results regarding the
stability of the FC and ZFC states. In the anomalous peak effect region, the FC state
has been observed to be unstable against current annealing. The critical current $I_c$
is higher (upto a factor of 6) in the virgin FC state than in FC states annealed by a
current or in the ZFC state\cite{hender}. On the other hand, using a slow ramp rate for
$I$, the FC and ZFC state $I_{c}$ was found to be same below and inside the peak effect
regime\cite{xiao}. This was interepreted as a signature of the metastability of the FC
state. Also, the susceptibility of the FC state can be switched to that of the ZFC state
by applying an ac pulse \cite{banerjee,sergio}. Moreover, recent scanning Hall probe
measurements have also shown substantial vortex rearragement in FC states for $I<I_{c}$
\cite{march}. One approach accounts for these experimental observations through the role
of surface barriers \cite{paltiel}. These differing pictures place in the focus the
behavior of vortices in the FC and ZFC states in the presence of a transport current.

In this paper we use numerical simulations to study the transport of differently prepared
vortex systems. Our first main result is that, in contrast to the temperature cycling, if
the vortex system is cycled with a transport current, then the ZFC state is more stable
than the FC state. In particular, the ZFC state can be reached from the FC state by
current cycling, but not vice versa. Our second main result is that, with increasing
currents, all vortex states, FC, ZFC, or prepared in any intermediate way, acquire an 
unique {\it density of defects at depinning}.

We consider vortices in a 2D plane perpendicular to the magnetic field ${\bf
B}=B\hat{\bf z}$ of a bulk superconductor, with $B=n_{v}\phi_{0}$ where $n_{v}$ is
the vortex density, and $\phi_{0}$ is the flux quantum. Within the London
approximation we treat the vortices as particles, with dynamics governed by an
overdamped equation of motion, 
\begin{eqnarray} \eta\frac{d{\bf r}_{i}}{dt} =
-\sum_{j\neq i} \nabla U^{v}({\bf r}_{i}-{\bf r}_{j}) - \sum_{k} \nabla U^{p}({\bf
r}_{i}-{\bf R}_{k}) + \nonumber \\ 
{\bf F}_{ext} + {\bm \zeta}_{i}(t) . \nonumber
\end{eqnarray} 
Here $\eta$ is the flux-flow viscosity, and the first term represents
the inter-vortex interaction given by the potential
$U^{v}(r)=\frac{\phi_{0}^{2}}{8\pi^{2}\lambda^{2}} K_{0}(\tilde{r}/\lambda)$, where
$K_0$ is the zeroth-order Bessel function, and $\tilde{r} = (r^{2}+2\xi^{2})^{1/2}$
with $\lambda$ and $\xi$ as penetration depth and coherence length of the
superconductor, respectively \cite{clem}.  The second term is the attractive
interaction with parabolic potential wells
$U^{p}(r)=U_{0}(\frac{r^{2}}{r_{p}^{2}}-1)$ for $r < r_{p}$, and 0 otherwise,
centered at the random ${\bf R}_{k}$ locations.  The third term
$F_{ext}=\frac{1}{c}{\bf J}\times \phi_{0}\hat{\bf z}$ represents the Lorentz force
due to the transport current density ${\bf J}$. The last term is the thermal noise
(Langevin term) with the moments $\langle\zeta_{i,p}(t)\rangle\!=\!0$, and
$\langle\zeta_{i,p}(t)\zeta_{j,p'}(t')\rangle\!=\!
2k_{B}T\eta\delta_{ij}\delta_{pp'}\delta(t-t')$, where $T$ is the temperature,
$k_{B}$ is Boltzmann constant, and $p,p'\!=\!x,\!y$.  We worked with reduced
variables: all distances are in units of $\lambda_{0}\!=\!\lambda(T\!=\!0)$, the
current density $J$ in units of $cf_{0}/\phi_{0}$, and $T$ in units of
$\lambda_{0}f_{0}/k_{B}$ where
$f_{0}\!=\!\frac{\phi_{0}^{2}}{8\pi^{2}\lambda_{0}^{3}}$. The time $t$ is in units
of $\eta\lambda_{0}/f_{0}$. This model is expected to reasonably capture the physics
of stiff vortex lines, like those in Nb and NbSe$_2$, while in strongly
layered superconductors, the soft interlayer couplings introduce additional degrees of 
freedom.

To simulate the ZFC state, we confine the quenched disorder to a central region of the
simulation box, while leaving the outer region defect free as illustrated in Fig.~1. The
disordered region defines the extent of the superconductor and the disorder free region
simulates the free-space. Periodic boundary conditions connect the edges of the
simulation box. We employ a box of size $40\lambda_{0}\times 24\lambda_{0}$, with the
disordered region occupying $28\lambda_{0}\times 24\lambda_{0}$. We used a pin density
of 5.95/$\lambda^{2}_{0}$ with $\langle U_{0}\rangle\!=\!0.03$, and
$r_{p}=0.1\lambda_{0}$.  The results presented below are for $N_{v}=1148$ vortices in
the simulation box which corresponds to $B\approx 800 G$. Typically, we used
$5\times10^{4}$ time steps for thermal equilibriation. For $V$-$I$ characteristics,
$2\times10^4$ time steps was used to equilibriate before averaging over a similar time
scale. The real space configuration of vortices is characterized by the defect density
$n_{d}\!=\!N_{d}/S$ where $N_d$ is the number of voritces which are non six-fold
co-ordinated and $S$ is the area (in units of $\lambda_{0}^{2}$) of the disordered
region\cite{comment1}. $N_{d}$ is counted by Delaunay triangulation of the real space 
position of vortices. The $V$-$I$ characteristics are calculated by varying ${\bf
F}_{ext}=F_{ext}\hat{\bf x}$ ($F_{ext} \propto I$, the current)  and calculating the
average velocity ($\propto V$, the voltage) ${\bf
v}=\frac{1}{N_{v}}\sum_{i}^{N_{v}}\dot{\bf r}_{i}$ along the $x$-axis.

We first consider the effect of cycling the temperature $T$. The ZFC state at
$T\!=\!0$ is created by positioning vortices in the defect free region and allowing
them to move into the disordered region by their own dynamics. The FC state is created
by the thermal cycle of the ZFC state by first heating the system deep into the liquid
phase, and subsequently cooling it back to $T\!=\!0$. Fig.~1 shows the behaviour of
the defect density $n_d$ throughout this thermal cycle. The ZFC state at $T\!=\!0$ is
formed by vortices diffusing through the free edges into the disordered region. With
decreasing velocity of the diffusion front, plastic deformation induces a larger
density of defects. These defects are non-equilibrium and is supported by a gradient
($\propto\!\langle U^{p}\rangle$) in vortex density\cite{bean}. On increasing $T$, the
gradient, and subsequently $n_d(T)$, decreases due to thermal activation. Across
$T\!=\!0.012\!=\!T_{m}$, the melting temperature, $n_d(T)$ increases rapidly as the
system goes into the liquid phase. Compared to the ZFC state, the FC state has a lower
$n_d(T\!=\!0)$ as it is cooled from the homogeneous liquid phase that has no
macrosopic density gradients. The FC state shows no appreciable change in $n_d$ on
repeated thermal cycling. The inset in Fig.~1 shows the hysteresis in $n_d$ on
crossing $T_m$, which is evidence for the melting transition into the liquid phase
being first order. A very slow cooling rate is used in the inset compared to the main
panel to more accurately observe the hysteresis across $T_{m}$.

\begin{figure}[hbt]
\includegraphics[width=230pt]{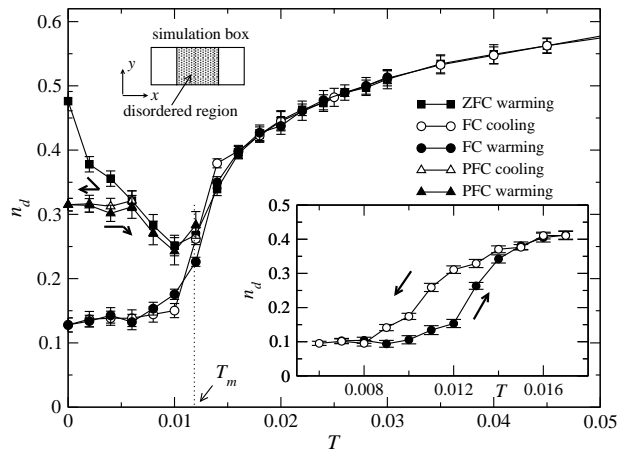}
\caption{\label{thyst} Plot of $n_{d}(T)$ for the thermal cycling starting with the ZFC
state at $T\!=\!0$. Also shown is the partial stability of the PFC state, created by 
cooling from $T_{PFC}\!=\!0.006$. The direction in which $T$ is changed for creating 
the PFC state is shown by thick arrows. The simulation geometry is shown on the upper 
left corner of the main panel. Inset : hysteresis in $n_d$ across $T_{m}$ during the 
FC cooling and warming procedure.}
\vspace{-0.3cm}
\end{figure}

A partial thermal cycling of the system, consisting of a warming step to a temperature
$T_{PFC}$, e.g. $T_{PFC}\!=\!0.006\!=\!\frac{T_{m}}{2}$, and a subsequent cooling step to
$T\!=\!0$ generates a partially field-cooled (PFC) state (see Fig.~1). The PFC state is
less disordered than the ZFC state, and is stable to thermal cycling between $0$ and
$T_{PFC}$. For $T\!>\!T_{PFC}$, the PFC state follows the same path as the virgin ZFC
state. This repeated warming and cooling allows one to access all the other PFC
configurations at $T\!=\!0$. The behaviour of $n_{d}(T)$ is in consonance with the
experimental measurement of magnetisation $M(T)$ for $T\!<\!T_{irr}$\cite{bednorz}.

Next we consider the effect of current $I$ on various configurations. As shown in
Ref.\cite{kosh}, effects of the disorder on the moving vortex state can be
characterized by a ``shaking temperature'' $T_{sh}$. $T_{sh}\propto v^{-1}$, where
$v\propto I$ is the average vortex velocity. At large $v$, a moving lattice is formed
which undergoes an order to disorder transition as $I$ is decreased to zero across the
depinning current $I_c$. Therefore, besides the above described thermal cycling, a
disordered vortex state can be prepared by current cycling as well. As will be shown
below, the resulting configuration is the most stable vortex state against repeated
cycling of $I$.  We call the resulting static configuration the current hardened (CH)  
state, in analogy to ``strain hardening'', in which materials are cycled with
increasing stress to increase the density of dislocations.

Fig.~2 shows $V(I)$ and $n_{d}(I)$ curves at $T\!=\!0$ for differently prepared states. We
identify three current regimes : (1) a pinned regime $I\!<\!I_{c}$, with $V\!=\!0$ and
$n_{d}(I)$ determined by the mode of preparation; (2) a plastic regime
$I_{c}\!<\!I\!<\!I_{m}$, with non-linear $V(I)$ and the {\it same} $n_d(I)$ for all
differently prepared states; and (3) a flowing-lattice regime $I\!>\!I_{m}$, with linear
$V(I)$ and $n_{d}(I)$ substantially smaller. The small difference in $n_{d}$ in the
flowing-lattice regime, as seen in Fig.~2, is due to the formation of a grain boundary at
the free edges of the disorder region (see also Fig.~4(c)). Remarkably, $V(I)$ curves and
the depinning current $I_c$ are independent of the mode of preparation of the 
system\cite{comment4}, in agreement with the experiments with slow ramping of 
$I$\cite{xiao}. These characteristics persist at finite $T$ as well. The 
$n_{d}(I)$ for $I\!<\!I_c$, on the other hand, depends on the mode of preparation.

\begin{figure}[ht]
\includegraphics[width=230pt]{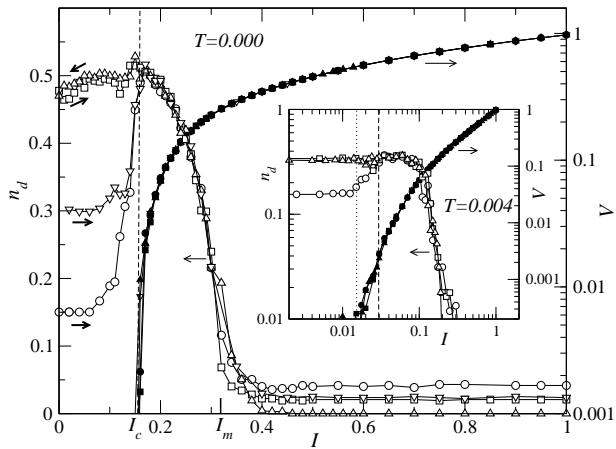} 
\caption{\label{iv}
$V(I)$ (filled symbols), and $n_{d}(I)$ (open symbols) at $T\!=\!0$ for the FC 
state ($\circ$), PFC state ($\bigtriangledown$), ZFC state ($\Box$), and CH
state ($\triangle$). The direction in which current is changed is shown by thick arrows. 
Inset : same as the main panel for $T\!=\!0.004$. The dashed and the dotted lines mark
$I_{irr}$ and $I_{c}$, respectively.}
\vspace{-0.3cm}
\end{figure}

Strikingly, the defect density $n_{d}(I)$ at the depinning current $I_{c}$ assumes
the {\it same} value $n_{d}^{c}\!\approx 0.51\!$ for the FC, ZFC, PFC, and CH
states, as seen in Fig.~2. In other words, $n_{d}^{c}$ is {\it independent of the
mode of the preparation of the vortex state}.  The history independent unique value
of $n_{d}^{c}$ implies that the characteristics and dynamics of all vortex states
converge as the current approaches $I_{c}$. This idea is substantiated by studying
the real space dynamics at depinning. As is known, vortices depin by
forming few channels across the system \cite{moon}. We find that, while in
differently prepared states the channels are nucleated at different locations along
the entry edge, in the bulk the channels converge quickly to a universal pattern. We
also find the average transverse wandering of a channel is of the order of $\sim
6$-$7$ lattice constants, and is same for all states at $I_{c}$. This implies that
the average displacement of the vortices at $I_{c}$ is the same for all states, thus
providing an underlying physical picture for preparation independence of
$n_{d}^{c}$. Further studies show that the value of $n_{d}^{c}$ depends on $T$,
$n_{v}$, and the disorder strength.

Next, we study the stability of the different vortex states against cycling with
$I$. The virgin FC state has a defect density $n_{d,FC}(0)\!<\!n_{d}^{c}$.  
Therefore, the density of defects in the virgin FC state increases from
$n_{d,FC}(0)$ to $n_{d}^{c}$ as the $I$ is increased from 0 to $I_{c}$, as shown in
Fig.~2. Furthermore, decreasing current from $I_{1}\!<\!I_{c}$, $n_{d,FC}(I)$ does
not decrease to its virgin value, but stays approximately at $n_{d}(I_{1})$. In
particular, cycling $I$ between 0 and $I\!\geq\!I_{c}$ leaves $n_{d,FC}(I)\approx
n_{d}^{c}$ for $I\!<\!I_{c}$. At the same time, $n_{d,ZFC}(0)\approx n_{d}^c$, and
cycling $I$ does not affect this value. Therefore, the FC state, and its defect
density $n_{d,FC}$, can be transformed into the ZFC state and $n_{d,ZFC}$ by cycling
$I$ between 0 and $I_c$. The reverse is not true, {\it i.e.}, {\em the FC state
cannot be reached from any state by cycling current}. Therefore, in contrast to
thermal cycling, we find that the FC state is metastable and the ZFC state is stable
against current cycling. Fig.~2 further illustrates that the CH state and its defect
density $n_{d,CH}(I)$ evolve into the ZFC state when the current is decreased from a
large value to 0. The same holds for the PFC states, as $n_{d,PFC}(I)$ also evolves
into $n_{d,ZFC}(0)$.

Since for $I\!>\!I_{c}$ $n_{d}(I)$ is approximately the same for all states, metastable
configurations and the irreversible behavior of $n_{d}(I)$ emerge only below $I_{irr}$. For
$T\!=\!0$, $I_{irr}\!=\!I_{c}$, but at finite temperatures these two current values
separate, as evident for $T\!=\!0.004$, shown in the inset of Fig.~2. At finite $T$,
$n_{d}(I) \approx n_{d}^{c}$ at a current $I_{irr}\!>\!I_{c}$. This is possibly due to
thermally activated vortex creep, which induces vortex displacements already for
$I\!<\!I_{irr}$.  The metastability of the FC state persists at finite $T$, and we believe
is a generic phenomenon which brings out a fundamental difference in the nature of FC and
ZFC states. The ZFC and CH states show similar behaviour throughout the studied current and
temperature ranges.

\begin{figure}[hbt]
\includegraphics[width=230pt]{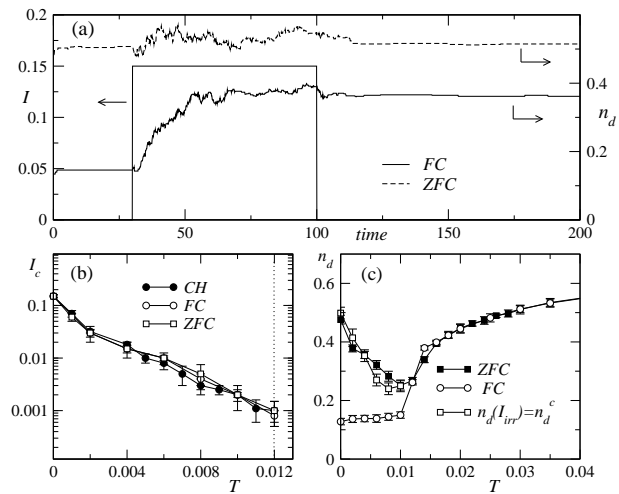}
\caption{\label{pulse}(a) The time evolution of $n_{d}$ at $T=0.0$ for a pulse of current
$I=0.15<I_{c}$ in the ZFC and the FC state. (b) Plot of $I_{c}(T)$ obtained from FC, ZFC 
and CH states. (c) Shows $n_{d}^{c}(T)$ obtained from $V(I)$ curves along with $n_d$ for
ZFC warming and FC cooling superimposed from Fig.~1.}
\vspace{-0.3cm}
\end{figure}

Further evidence for metastability of the FC state is obtained by applying a current pulse
$I=0.15\!<\!I_{c}$ to differently prepared vortex states, and observing the temporal
evolution of $n_{d}$, as illustrated in Fig.~3(a). The application of a current pulse
evolves the FC state into PFC states with an intermediate defect density
$n_{d,FC}\!<\!n_{d,PFC}\!<\!n_{d,ZFC}$. The defect density remains at that elevated level
even after the current drops to zero. In particular, a large enough current pulse evolves
the FC state into the ZFC state. This transformation of the FC state to the ZFC state by a
current pulse is in close analogy to experimental observations \cite{banerjee}.  In
Ref.\cite{banerjee}, the $I_c$ is also found to change during the transformation of the FC
state to ZFC state, possibly due to the influence of surface barrier\cite{xiao1}. We
emphasize that the transformation of the FC configuration to ZFC configuration as found
here is a bulk characteristic, and is not influenced by the free edges of the disordered
region. This was verified in simulation without free edges\cite{comment2}. The FC state was
again found to be metastable below $I_{irr}$, and is transformed into CH configuration on
increasing $I$ above $I_{irr}$.

In Fig.~3(b) we plot $I_{c}(T)$ for a range of $T$ at which simulations were carried out.
The $I_{c}$ follows an $\exp(-\frac{T}{T_{0}})$ form, confirming that thermally activated
motion dominates the dynamics at $I_{c}$. Fig.~3(c) shows the temperature dependence of
$n_{d}^{c}(T)$, as obtained from $V(I)$ curves. The $n_d(T)$ values for the FC and ZFC
states are superimposed from Fig.~1. $n_{d}^{c}(T)$ closely follows $n_{d,ZFC}(T)$,
obtained by warming the ZFC state, thus corroborating the view that the ZFC state is
essentially the frozen equivalent of the critical state at $I_{c}$. In other words, 
the channels formed by the penetrating front of vortices during the formation of the 
virgn ZFC state is frozen. Any subsequent application of $I$ essentially uses these 
pre-formed channels which explains why ZFC state shows no substantial change in 
$n_d$. In contrast, in the virgin FC state no such channels exists. On first 
passage of $I$, the resulting build up of stress at depinning leads to channel 
formation which increases $n_d$. Any further cycling of $I$ essentially uses the 
already formed channels. This is a plausible explanation for the metastability of 
the FC state on cycling $I$.

\begin{figure}[htb]
\includegraphics[width=230pt]{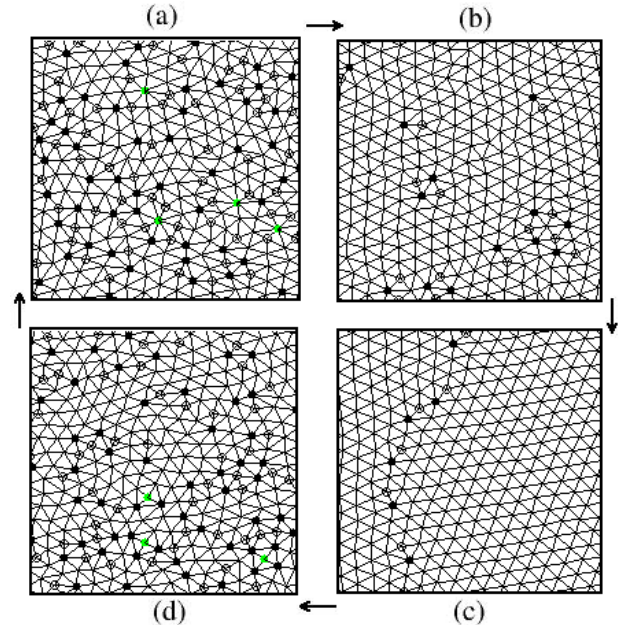}
\caption{\label{config}The real space configuration at $T\!=\!0$ during a composite $T$-$I$ 
cycle : (a) ZFC state, (b) FC state, (c) flowing lattice state, 
(d) CH state. The arrows show the cycling path (see text for details). Note the close 
similarity between (a) and (d). The symbols ($\circ$) and ($\bullet$) denote vortices with 
5 and 7 neighbours, respectively.}
\end{figure}

The configurational changes are illustrated in Fig.~4 for various states, during a
composite $T$-$I$ cycle. Starting with the ZFC state, shown in Fig.~4(a), we obtain
the FC state (Fig.~4(b)) by cycling $T$ from $0$ to $T\!>\!T_{m}$. The FC state is
then driven with $I\!>\!I_{m}$ to obtain Fig.~4(c). And finally, decreasing $I$ to
0, the CH state is obtained in Fig.~4(d), which is indistinguishable from the ZFC
state in Fig.~4(a). {\it Thus a composite $T-I$ cycle allows one to go from the ZFC
state to the FC state and vice-versa}.

In conclusion, we have shown that the ZFC state is stable against current cycling in
contrast to the FC state which is found to be metastable. This behavior is
opposite to the thermal cycling where the FC state is stable. We also find that the 
defect density $n_{d}$ for any initial state converges to the same value at $I_{c}$.

We thank Helmut Katzgraber for critical reading of the manuscript. M.C. also acknowledges 
useful discussions with E. Zeldov and A. K. Grover during the course of the work. This work 
was supported by NSF-DMR 9985978.


\begin{references}
\bibitem{comment1} We choose to define defect density rather than defect fraction because 
the number of vortices within the disordered region is not fixed but fluctuates with time.
\bibitem{comment4} We find different $I_c$'s on repeated current cycling due to  
metastable states present during plastic depinning. But the distribution width of $I_c$'s 
is same for any starting configuration, viz. FC, ZFC, PFC and CH, and is not more than 
2\% of the average $I_c$.
\bibitem{comment2} Since the ZFC state cannot be simulated without free edges, we 
compared only the FC and the CH states.
\end{references}
\end{document}